\documentclass[12pt,a4paper]{article}
\usepackage{a4wide}
\usepackage{amsmath}
\usepackage{amssymb}
\usepackage{epsfig}

\def\lsim{\raise0.3ex\hbox{$<$\kern-0.75em\raise-1.1ex%
\hbox{$\sim$}}}
\def\gsim{\raise0.3ex\hbox{$>$\kern-0.75em\raise-1.1ex%
\hbox{$\sim$}}}

\begin{document}

\begin{titlepage}
  \title{Supernova Neutrino Oscillations\footnote{Contribution to the
      Proceedings of the Nobel Symposium 129 ``Neutrino Physics,''
      19--24 August 2004, Enk\"oping, Sweden, to be published in
      Physics Scripta (Topical Issues), eds.\ L.~Bergstr\"om,
      O.~Botner, P.~Carlson, P.~O.~Hulth and T.~Ohlsson}}

\author{Georg G.~Raffelt\\ [4mm]
{\it {\footnotesize 
Max-Planck-Institut f\"ur Physik (Werner-Heisenberg-Institut)}}\\
{\it{\footnotesize F\"ohringer Ring 6, 80805 M\"unchen, Germany}}
\date{7 January 2005}
}
\maketitle 

\begin{abstract}
  Observing a high-statistics neutrino signal from a galactic
  supernova (SN) would allow one to test the standard delayed
  explosion scenario and may allow one to distinguish between the
  normal and inverted neutrino mass ordering due to the effects of
  flavor oscillations in the SN envelope. One may even observe a
  signature of SN shock-wave propagation in the detailed
  time-evolution of the neutrino spectra.  A clear identification of
  flavor oscillation effects in a water Cherenkov detector probably
  requires a megatonne-class experiment.
\end{abstract}

\bigskip


\end{titlepage}

%%%%%%%%%%%%%%%%%%%%%%%%%%%%%%%%%%%%%%%%%%%%%%%%%%%%%%%%%%%%%%%%%%%%%%
\section{Introduction}
%%%%%%%%%%%%%%%%%%%%%%%%%%%%%%%%%%%%%%%%%%%%%%%%%%%%%%%%%%%%%%%%%%%%%%

While galactic supernovae are rare, the proliferation of existing or
proposed large neutrino detectors has considerably increased the
confidence that a high-statistics supernova (SN) neutrino signal will
eventually be observed. The scientific harvest would be immense. Most
importantly for particle physics, the detailed features of the
neutrino signal may reveal the nature of the neutrino mass ordering
that is extremely difficult to determine experimentally
(e.g.~Ref.~\cite{Huber:2004ug} and references). On the other hand, a
detailed measurement of the neutrino signal from a galactic SN could
yield important clues on the SN explosion mechanism. Neutrinos
undoubtedly play a crucial role for the SN dynamics and in particular
neutrino energy deposition behind the SN shock is able to initiate and
power the SN explosion~\cite{Wi85,BW85}. However, it is still unclear
whether this energy deposition is sufficiently strong because current
state-of-the-art models still have problems to produce robust
explosions~\cite{BRJK03}.  Empirical constraints on the physics deep
inside the SN core would be extremely useful because neutrinos are the
only way for a direct access~\cite{Totani:1997vj} besides
gravitational waves~\cite{Mueller:2003fs}.

The neutrinos emitted by the collapsed SN core will pass through the
mantle and envelope of the progenitor star and on the way encounter a
vast range of matter densities $\rho$ from nearly nuclear at the
neutrinosphere to that of interstellar space. The Wolfenstein
effect~\cite{Wolfenstein:1977ue} causes a resonance in neutrino
oscillations~\cite{Mikheev:gs} when $\Delta
m^2_\nu\cos2\theta/2E_\nu=\pm\sqrt{2}G_{\rm F}Y_{\rm e}\rho$, where
the plus and minus sign refers to neutrinos $\nu$ and antineutrinos
$\bar\nu$, respectively. Therefore, depending on the sign of $\Delta
m^2_\nu$, the resonance occurs in the $\nu$ or $\bar\nu$
channel~\cite{Langacker}. For the ``solar'' neutrino mass-squared
difference of $\Delta m^2_{21}\approx 79~{\rm
meV}^2$~\cite{Araki:2004mb} one refers to the ``L-resonance'' (low
density) while for the ``atmospheric'' one of $|\Delta
m^2_{32}|\approx 2300~{\rm meV}^2$~\cite{Maltoni:2004ei} to the
``H-resonance'' (high density).  The resonance is particularly
important for 13-oscillations because the 13-mixing angle is known to
be small so that the classic MSW enhancement of flavor conversion in
an adiabatic density gradient is a crucial
feature~\cite{Dighe:1999bi}.

Effects of neutrino flavor oscillations will be observable only if the
fluxes and/or spectra emitted at the source depend on the neutrino
species. Therefore, I will summarize in Sec.~\ref{sec:sourcespectra}
the current understanding of SNe as flavor-dependent neutrino
sources. In Sec.~\ref{sec:flavoroscillations} the effect of neutrino
flavor oscillations in different mixing scenarios will be summarized.
Sec.~\ref{sec:observations} is devoted to experimentally observable
signatures before turning to conclusions in
Sec.~\ref{sec:conclusions}.

%%%%%%%%%%%%%%%%%%%%%%%%%%%%%%%%%%%%%%%%%%%%%%%%%%%%%%%%%%%%%%%%%%%%%%
\section{Core-Collapse Supernovae as Neutrino Sources}
%%%%%%%%%%%%%%%%%%%%%%%%%%%%%%%%%%%%%%%%%%%%%%%%%%%%%%%%%%%%%%%%%%%%%%

\label{sec:sourcespectra}

The collapsed core of a SN is essentially a blackbody source for
neutrinos of all flavors with a temperature of several MeV,
corresponding to the temperature of the medium near the proto-neutron
star's surface. In detail, however, the fluxes and spectra differ
between the different species because of their different interaction
channels. Electron neutrinos and anti-neutrinos interact only by
charged-current processes on nucleons while the other species interact
primarily by neutral-current reactions because the relevant energies
and densities are too low to produce muons or tau-leptons. Therefore,
$\nu_\mu$, $\bar\nu_\mu$, $\nu_\tau$, and $\bar\nu_\tau$ will be
emitted with identical fluxes and spectra and are thus collectively
referred to as $\nu_x$.

\begin{figure}[ht]
\begin{center}
\epsfig{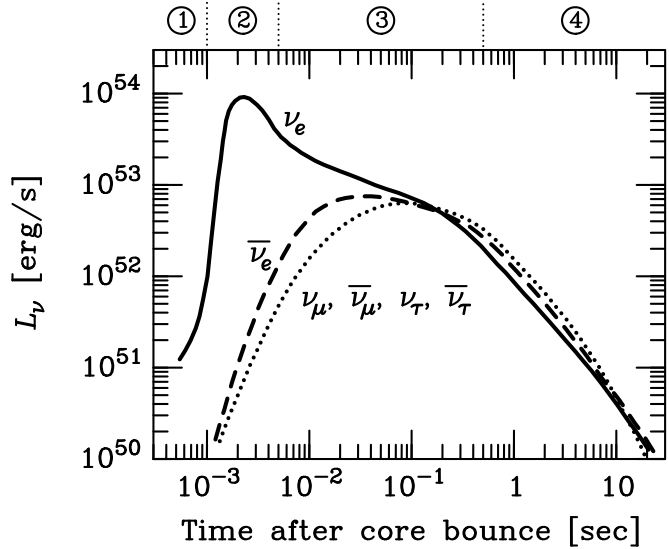}
\end{center}
\caption{Schematic SN neutrino ``light curves'' for different 
neutrino flavors~\cite{Raffelt:1996wa}.}
\label{fig:lightcurve}
\end{figure}

The standard delayed-explosion scenario of SN evolution and thus the
neutrino emission is characterized by four distinct phases,
schematically indicated in Fig.~\ref{fig:lightcurve}: (1)~Core
collapse and bounce.  (2)~Shock propagation and
breakout. (3)~Accretion and mantle cooling while the shock wave
stagnates.  (4)~Kelvin-Helmholtz cooling of the neutron star after the
explosion.  Actually, for a very close star the neutrinos emitted
during the silicon-burning phase for the last few days before collapse
may be observable~\cite{Odrzywolek:2003vn}, a possibility that I will
not further consider here.

The $\nu_e$ ``light curve'' shows a distinct peak, the ``prompt
deleptonization burst,'' that occurs when the shock-wave breaks
through the neutrino sphere, dissociates iron and thus allows for the
quick $e^-+p\to n+\nu_e$ conversion and thus for the prompt
deleptonization of the outermost core layers. The flux and spectral
characteristics of this burst are probably the least model-dependent
features of SN neutrino
emission~\cite{Takahashi:2003rn,Kachelriess:2004ds}. During the phase
of shock-wave stagnation that may last for several hundred
milliseconds, the neutrino emission is largely powered by the
accretion of matter. After the shock wave has been rejuvenated by
neutrino energy deposition and has ejected the stellar mantle and
envelope, the neutron star remnant will cool, i.e.\ the neutrino
emission is powered by energy stored deep in the core that diffuses to
the surface on a time scale of several seconds.

In the literature it was often assumed that the neutrino emission
during the accretion and Kelvin-Helmholtz phases was characterized by
an approximate equipartition of the flavor-dependent neutrino
luminosities $L_{\nu_e}\approx L_{\bar\nu_e}\approx L_{\nu_x}$ and by
a strong hierarchy of average energies $\langle
E_{\nu_e}\rangle\ll\langle E_{\bar\nu_e}\rangle\ll \langle
E_{\nu_x}\rangle$. This behavior is borne out, for example, by the
numerical simulations of the Livermore group~\cite{Totani:1997vj}.
However, traditional treatments of neutrino transport involve a number
of physical and numerical simplifications that were justified for the
purposes of those calculations but are not accurate enough to judge
the flavor dependence of the neutrino fluxes and spectra. In a series
of papers we have studied this problem and have concluded that the
differences between $\langle E_{\bar\nu_e}\rangle$ and $\langle
E_{\bar\nu_x}\rangle$ are probably not more than about 20\% or
less~\cite{Raffelt:2001kv,Buras:2002wt,Keil:2002in,Raffelt:2003en}.
On the other hand, the equipartition of energy among the flavors is
certainly not exact---the luminosities can differ perhaps by up to a
factor of~2. Moreover, it appears that during the accretion phase
$L_{\bar\nu_x}<L_{\bar\nu_e}$ while $L_{\bar\nu_e}<L_{\bar\nu_x}$
during the cooling phase~\cite{Keil:2002in,Raffelt:2003en}.

While several recent numerical simulations with a modern treatment of
neutrino transfer confirm this picture for the accretion phase, no
up-to-date simulations have treated the long-term evolution of the
neutrino signal during the cooling phase. Since current numerical
models to not produce robust explosions and since a full Boltzmann
treatment of neutrino transport is very CPU intensive, numerical
studies of the flavor-dependence of the late neutrino signal are not
available at the present time. Therefore, forecasting possible effects
of neutrino oscillations in the signal from a future galactic SN is
unavoidably schematic and must rely on generic assumptions about the
flavor-dependent neutrino fluxes and spectra.

%%%%%%%%%%%%%%%%%%%%%%%%%%%%%%%%%%%%%%%%%%%%%%%%%%%%%%%%%%%%%%%%%%%%%%
\section{Impact of Flavor Oscillations}
%%%%%%%%%%%%%%%%%%%%%%%%%%%%%%%%%%%%%%%%%%%%%%%%%%%%%%%%%%%%%%%%%%%%%%

\label{sec:flavoroscillations}

The neutrino or anti-neutrino fluxes arriving at Earth from a SN are
determined by the primary spectra at the source as well as neutrino
oscillation effects. For $\nu_e$, the flux arriving at Earth may be
written in terms of the energy-dependent ``survival probability''
$p(E)$ as
\begin{equation}
F_{\nu_e}(E) = p(E) F_{\nu_e}^0(E) + [1-p(E)] F_{\nu_x}^0(E)\,,
\label{pbar-def}
\end{equation}
where the superscript zero denotes the primary fluxes and $\nu_x$
stands for either $\nu_\mu$ or $\nu_\tau$. An analogous expression
pertains to $\bar\nu_e$ with the survival probability $\bar p(E)$.

When neutrinos propagate through the SN mantle and envelope, the MSW
resonance corresponding to the solar mixing angle is adiabatic and is
always in the neutrino channel. On the other hand, the H-resonance,
corresponding to $13$-oscillations, is in the neutrino channel for
normal neutrino mass ordering while it is in the anti-neutrino channel
in the inverted case. This resonance is adiabatic for all relevant
energies if the mixing angle $\theta_{13}$ is ``large'' in the sense
of $\sin^2\theta_{13}\,\gsim\,10^{-3}$.  It is non-adiabatic if the
mixing angle is ``small'' in the sense of
$\sin^2\theta_{13}\,\lsim\,10^{-5}$.  For intermediate values, the
adiabaticity depends on energy and the situation is more complicated.
The survival probabilities thus depend on the magnitude of the
13-mixing angle and on the nature of the mass ordering so that, in
principle, the observation of a SN neutrino signal can distinguish
between different mixing scenarios~\cite{Dighe:1999bi}.
Table~\ref{tab:survival} summarizes the survival probabilities for
different mixing scenarios where $\theta_\odot$ refers to the
``solar'' mixing angle.

\begin{table}
\begin{center}
\begin{tabular}{llllllll}
\hline
Scenario&Hierarchy& $\sin^2\theta_{13}$ & $p$ & $\bar{p}$&
Earth effects&Shock wave&$\nu_e$ burst\\
\hline
A & Normal &${\gsim}\,10^{-3}$  & 0  & $\cos^2\theta_\odot$&
$\bar\nu_e$& $\nu_e$&absent \\
B & Inverted &  $\gsim\, 10^{-3}$ &  $\sin^2\theta_\odot$ &  0&
$\nu_e$& $\bar\nu_e$&present\\
C & Any & $\lsim\, 10^{-5}$  & $\sin^2\theta_\odot$ 
&  $\cos^2\theta_\odot$ &
$\nu_e$ and $\bar\nu_e$&---&present\\
\hline
\end{tabular}
\caption{Survival probabilities for neutrinos, $p$, and antineutrinos,
  $\bar{p}$, in various mixing scenarios.  The channels where one
  expects Earth effects, shock-wave propagation effects, and where the
  full $\nu_e$ burst is present or absent are indicated.}
\label{tab:survival}
\end{center} 
\end{table}

The indicated survival probabilities pertain to a situation where
neutrino oscillations occur only in the SN mantle and envelope.  If
the SN is shadowed by the Earth for the relevant detector, additional
modifications of the survival probabilities arise due to Earth matter
effects, causing an energy-dependent modulation of $p(E)$ or $\bar
p(E)$. Table~\ref{tab:survival} indicates the channel where Earth
effects would arise for different mixing scenarios.

The SN shock wave will pass the density region corresponding to the
H-resonance a few seconds after core bounce, breaking the adiabaticity
of the 13-resonance. The resulting transient modification of the
survival probabilities may be observable~\cite{Schirato:2002tg}.  They
arise in the channels indicated in Table~\ref{tab:survival}.

%%%%%%%%%%%%%%%%%%%%%%%%%%%%%%%%%%%%%%%%%%%%%%%%%%%%%%%%%%%%%%%%%%%%%%
\section{Experimental Signatures}
%%%%%%%%%%%%%%%%%%%%%%%%%%%%%%%%%%%%%%%%%%%%%%%%%%%%%%%%%%%%%%%%%%%%%%

\label{sec:observations}

The most distinct effects of flavor oscillations arise in the $\nu_e$
channel because it sports the prompt deleptonization burst and even
during the accretion and cooling phases, the expected spectral
differences are much larger between $\nu_e$ and $\nu_x$ than between
$\bar\nu_e$ and $\bar\nu_x$. On the other hand, the existing and
realistically expected large detectors, notably Super-Kamiokande,
IceCube, and a future megatonne-class water Cherenkov detector, are
mostly sensitive to the inverse beta decay $\bar\nu_e+p\to n+e^+$,
although the $\nu_e$ channel can be measured by the elastic scattering
reaction $\nu_e+e^-\to e^-+\nu_e$. The largest existing $\nu_e$
detector is the Sudbury Neutrino Observatory (SNO) that will be shut
down after completing its solar neutrino programme and thus will not
be available for a long-term SN watch. Of course, if an efficient
$\nu_e$ detector such as as a large liquid Argon TPC should become
available, it would have unique capabilities for studying SN neutrino
oscillations~\cite{Gil-Botella:2003sz,Gil-Botella:2004bv}.

In a water Cherenkov detector, the $\nu_e$ signal can be identified by
the directionality of the $\nu_e$-$e^-$ elastic scattering
reaction. Moreover, if future water Cherenkov detectors are doped with
enough gadolinium to tag the neutron from $\bar\nu_e+p\to n+e^+$
\cite{Beacom:2003nk}, the identification of the prompt SN $\nu_e$
burst will become a realistic possibility. However, in
Super-Kamiokande the total number of expected events from the $\nu_e$
burst of a ``fiducial SN'' at a distance of 10~kpc is so small (about
a dozen) that a clear identification of the presence or absence of the
burst is not likely.  In a megatonne detector, on the other hand, the
full $\nu_e$ burst will produce about 200 events. Therefore, the time
structure of the SN signal during the first few tens of milliseconds
can provide a clean indication if the full $\nu_e$ burst is present or
absent~\cite{Kachelriess:2004ds} and therefore allows one to
distinguish between different mixing scenarios as indicated by the
last column of Table~\ref{tab:survival}. For example, if the mass
ordering is normal and the 13-mixing angle is large, the $\nu_e$ burst
will fully oscillate into $\nu_x$.  If the 13-mixing angle is measured
in the laboratory to be large, for example by one of the forthcoming
reactor experiments~\cite{Anderson:2004pk}, then one may distinguish
between the normal and inverted mass ordering.  On the other hand, if
the mixing scenario is independently identified by laboratory
experiments, these observations allow one to determine the distance to
the SN with a precision of about 5\%, even in the likely case that it
is optically obscured~\cite{Kachelriess:2004ds}.

A megatonne water Cherenkov detector is a realistic future possibility
in view of the world-wide drive towards precision long-baseline
oscillation experiments, but it is less clear if a gadolinium-doped
detector of such size will ever come into existence. Therefore, I now
turn to the main $\bar\nu_e$ detection channel of an ordinary water
Cherenkov detector. In this channel the main problem is that one can
not rely on theoretical predictions of the flavor dependence of the
source fluxes and spectra. Therefore, one has to rely on
model-independent signatures. A clear positive detection of such a
signature would indicate a specific mixing scenario while the absence
would be ambiguous: It could either imply a different mixing scenario
or could imply that the flavor-dependent source spectra and fluxes
were too similar to provide significant oscillation effects.

One unequivocal indication of oscillation effects would be the
energy-dependent modulation of the survival probability $\bar p(E)$
caused by Earth matter effects.  Even without an identification of the
SN location in the sky in the electromagnetic spectrum, its direction
can be established with sufficient accuracy by the $\nu_e$-$e^-$
elastic scattering signal
alone~\cite{Beacom:1998fj,Tomas:2003xn}. With a megatonne-class
detector, the modulation signal could be well established unless the
flavor-dependent flux differences are surprisingly
small~\cite{Dighe:2003jg,Dighe:2003vm}. The Earth effect would show up
in the $\bar\nu_e$ channel for the normal mass hierarchy, assuming
that $\theta_{13}$ is large (Table~\ref{tab:survival}).

Another possibility to establish the presence of Earth effects is to
use the signal from two detectors if one of them sees the SN shadowed
by the Earth and the other not. As one needs to compare the two
signals on the level of a few percent, sufficiently large detectors
are needed. The statistical fluctuations may be too large in
Super-Kamiokande so that, again, a megatonne-class detector is needed,
perhaps in combination with IceCube~\cite{Dighe:2003be}.  IceCube at
the South Pole and Hyper-Kamiokande in Japan would be geographically
complementary in that about 70\% of the sky would be seen directly in
one detector and shadowed by the Earth in the other.

At the time of the next galactic SN, the neutrino mixing parameters
may already be known from long-baseline laboratory experiments. In
that case it may become possible to use the SN neutrino signal in
reverse to perform a ``tomography'' of the Earth's density
profile~\cite{Lindner:2002wm}.

A few seconds after core bounce, the SN shock wave will pass the
density region in the stellar envelope corresponding to the
H-resonance and thus break the adiabaticity of 13-oscillations,
causing a transient modification of the survival probability and thus
a time-dependent signature in the neutrino
signal~\cite{Schirato:2002tg, Takahashi:2002yj, Lunardini:2003eh,
  Fogli:2003dw, Fogli:2004ff, Tomas:2004gr}. Probably the most
significant signature is the time variation of the average energy of
the detected positrons from the $\bar\nu_e+p\to n+e^+$ reaction. It
would show a characteristic dip when the shock wave passes, or a
double-dip feature if a reverse shock occurs~\cite{Tomas:2004gr}. This
signature may be observable in Super-Kamiokande and almost certainly
can be seen in a megatonne-class detector. Of course, apart from
identifying the neutrino mixing scenario, such observations would test
our theoretical understanding of the core-collapse SN phenomenon.

The shock-wave propagation signature arises in the complementary
channel to the Earth effects (Table~\ref{tab:survival}), i.e.\ it
shows up in the $\bar\nu_e$ channel for large $\theta_{13}$ and the
inverted mass hierarchy.  Therefore, unless the flavor-dependent
source spectra are unexpectedly similar, a megatonne water Cherenkov
detector is assured to see either shock-wave propagation effects or
Earth effects, although in the latter case it is required, of course,
that the SN is shadowed by the Earth.  Conversely, if neither Earth
effects nor shock-wave propagation effects are seen one has a serious
empirical handle on the flavor dependence of the source spectra in
that they must be unexpectedly similar.

%%%%%%%%%%%%%%%%%%%%%%%%%%%%%%%%%%%%%%%%%%%%%%%%%%%%%%%%%%%%%%%%%%%%%%
\section{Conclusions}
%%%%%%%%%%%%%%%%%%%%%%%%%%%%%%%%%%%%%%%%%%%%%%%%%%%%%%%%%%%%%%%%%%%%%%

\label{sec:conclusions}

Observing neutrinos from the next galactic SN would provide invaluable
information on the astrophysics of the core-collapse explosion
phenomenon and on neutrino mixing parameters. Existing or near-future
detectors such as Super-Kamiokande and IceCube are sufficient to
measure a precise $\bar\nu_e$ light curve and thus can establish, for
example, the duration of the accretion phase or perhaps detect
unexpected new features. However, these detectors are probably too
small to detect unambiguous evidence for neutrino oscillations. On the
other hand, a megatonne-class water Cherenkov detector is assured to
detect either Earth effects (assuming the SN is shadowed by the Earth)
or shock-wave propagation effects, depending on the neutrino mixing
scenario. Such observations will help to identify the neutrino mixing
scenario and test detailed aspects of theoretical SN physics.
Depending on the neutrino mixing information that will be established
by laboratory experiments alone and depending on theoretical progress
in our understanding of core-collapse SNe, one or the other aspect of
these observations will be of greater interest at the time of the next
galactic SN. Either way, the importance of such an observation as a
probe of neutrino physics and of the astrophysics of supernovae can
not be overstated.

%%%%%%%%%%%%%%%%%%%%%%%%%%%%%%%%%%%%%%%%%%%%%%%%%%%%%%%%%%%%%%%%%%%%%%
\section*{Acknowledgments} %%%%%%%%%%%%%%%%%%%%%%%%%%%%%%%%%%%%%%%%%%%
%%%%%%%%%%%%%%%%%%%%%%%%%%%%%%%%%%%%%%%%%%%%%%%%%%%%%%%%%%%%%%%%%%%%%%

This work was supported by the European Science Foundation (ESF) under
the Network Grant No.~86 Neutrino Astrophysics and by the Deutsche
Forschungsgemeinschaft (DFG) under grant No.~SFB-375.

%%%%%%%%%%%%%%%%%%%%%%%%%%%%%%%%%%%%%%%%%%%%%%%%%%%%%%%%%%%%%%%%%%%%%%
%% References %%%%%%%%%%%%%%%%%%%%%%%%%%%%%%%%%%%%%%%%%%%%%%%%%%%%%%%%
%%%%%%%%%%%%%%%%%%%%%%%%%%%%%%%%%%%%%%%%%%%%%%%%%%%%%%%%%%%%%%%%%%%%%%

\end{document}